\documentclass[a4]{article}
\usepackage{graphics}
\begin{document}

\title{Truncated first moment of the parton distribution - a modified approach}
\author{Dorota Kotlorz\footnote{Opole University of Technology,
 Division of Physics, Ozimska 75, 45-370 Opole, Poland, e-mail:
{\tt dstrozik@po.opole.pl}} \and Andrzej Kotlorz\footnote{Opole University
of Technology, Division of Mathematics, Luboszycka 3, 
45-036 Opole, Poland, e-mail: {\tt kotlorz@po.opole.pl}}}
\date{August 30, 2006}
\maketitle

\abstract{
We derive the LO DGLAP evolution equation for the full Mellin moments 
of the truncated at $x_0$ first moment of the nonsinglet parton 
distribution. This "moment of moment" approach allows to determine
the small-$x_0$ behaviour of the truncated first moment. We compare our 
predictions to results obtained from $x-$space solutions for parton 
distributions with use of the Chebyshev polynomial method and to solutions 
of the evolution equations for the truncated moments proposed by other 
authors. The comparison is performed for different input parametrisations
for $10^{-5}\leq x_0\leq 0.1$ and $1\leq Q^2\leq 100$ ${\rm GeV}^2$.
We give an example of application to the determination of the contribution
to the Bjorken Sum Rule.\\
PACS {12.38.Bx} {Perturbative calculations}, {11.55.Hx} {Sum rules}

\section{Introduction}
\label{intro}
Deep inelastic scattering (DIS) experiments provides knowledge 
about the internal structure of the nucleon. Measurements on 
proton, deuteron and neutron targets allow also verification of 
sum rules e.g. \cite{J1,J1a} - basic relations in QCD. Sum rules for 
unpolarised and polarised structure functions refer to their 
Mellin moments. Particularly important role in the experimental 
and theoretical QCD tests play first moments of the parton 
densities, which have physical interpretation and can be determined
from the data. From an experimental point of view, however, the 
accurate verification of sum rules is unreliable. Determination 
of the sum rules requires knowledge of the structure functions 
over the entire region of the Bjorken variable $x\in(0;1)$. The 
lowest limit of $x$ in present experiments is about $10^{-5}$ and 
the limit $x\rightarrow 0$, which means that the invariant
energy $W^2$ of the inelastic lepton-hadron scattering becomes 
infinite ($W^2=Q^2(1/x-1)$), will never be attained. Therefore 
it is very hopeful in the theoretical analysis to deal with 
truncated instead of with full moments of the structure functions.
This enables to avoid uncertainties from the unmeasurable 
$x\rightarrow 0$ region. The most familiar theoretical approach, which 
describes scaling violations of parton densities in perturbative QCD 
has been formulated by Dokshitzer, Gribov, Lipatov, Altarelli and Parisi
(DGLAP) \cite{J0}. Evolution equations for the 
truncated at low $x_0$ moments of parton distributions are however 
more complicated than in a case of the full moments. These are not 
diagonal and each $n$-th truncated moment couples to ($n+j$)-th 
($j\geq 0$) truncated moments \cite{J2}. For $n\geq 2$ the series 
of couplings to higher moments is very fast convergent. Even for 
small ($m=4$) number of terms in the expansion of the truncated 
counterpart of the anomalous dimension $G_n$, the higher moments 
can be calculated with excellent accuracy. First moment is more 
sensitive to the truncated point $x_0$ and the convergence of 
$G_n$ for $n=1$ is weaker than for the higher moments. Nevertheless, 
it has been shown in \cite{J3}, that for more terms of the $G_n$ 
expansion ($m\sim 30$), the uncertainty in the determination of 
the first moment at $x_0\leq 0.1$ and $Q^2=10$ ${\rm GeV}^2$ does 
not exceed $5\%$ independently on the input parametrisation. 
However, this increase with $m$ of the accuracy does not proceed 
infinitely. Numerical errors, which occur for larger $m$ 
(dependently on $x_0$) make further improvement of the precision 
impossible. It may be very useful to discuss methods of theoretical 
determination of truncated first moments of parton distributions 
because these predictions would be directly verified experimentally.
In this paper we present an approach, in which we compute the 
truncated at small $x_0$ first moments using the inverse Mellin 
transform of their full moments. In other words, our method is 
based on the solutions for the full $n$-th moments of the 
truncated first moment of the parton distribution. This "moment 
of moment" technique would be a complementary one to other
known methods in determination of sum rule contributions.

The content of this paper is as follows. In Section ~\ref{sec:2} 
we recall the ways of computing the truncated moments 
within DGLAP approximation. Thus Sect.~\ref{sec:2.1} contains 
a brief description of the Chebyshev-polynomial approach for 
$x$-space solutions of DGLAP evolution equations. In Sect.~\ref{sec:2.2} 
we recall the idea of DGLAP evolution equation for the truncated 
moments, which underlies our "moment of moment"  technique. This 
modified method is presented in Sect.~\ref{sec:2.3}. For simplicity 
we consider the quark nonsinglet LO evolution. In Section 
~\ref{sec:3}  we compare solutions for the truncated at small 
$x_0$ first moment of the nonsinglet structure function, obtained 
with use of these mentioned earlier approaches. As an example we explore
the Bjorken Sum Rule (BSR). In Section ~\ref{sec:4} we present predictions 
for the low-$x$ contribution to the BSR together with experimental 
constraints. Finally, in Section ~\ref{sec:5} we summarise our results.

\section{Determination of the truncated Mellin moments of parton 
distributions.}
\label{sec:2}
For (full) Mellin moments of parton distributions $p(x,Q^2)$
\begin{equation}\label{eq.1}
{\it M}\big[p\big](n,Q^2)=\int\limits_0^1 dx\: x^{n-1} p\,(x,Q^2)
\end{equation}
the DGLAP evolution equation can be solved analytically. This is 
because  in the moment space $n$ one obtains simple diagonalised 
differential equations. The only problem is the knowledge of the 
input parametrisation for the whole region $0\leq x\leq 1$, what 
is necessary in the determination of the initial moments 
${\it M}\big[p\big](n,Q^2=Q_0^2)$:
\begin{equation}\label{eq.2}
{\it M}\big[p\big](n,Q_0^2)=\int\limits_0^1 dx\: x^{n-1} p\,(x,Q_0^2).
\end{equation}
Using the truncated moments approach one can avoid the uncertainties 
from the region $x\rightarrow 0$, which will never be attained 
experimentally. 

The truncated at $x_0 $ Mellin moment of the parton distribution 
$p(x,Q^2)$ is defined as
\begin{equation}\label{eq.3}
{\it M}\big[p\big](x_0,n,Q^2)=\int\limits_{x_0}^1 dx\: x^{n-1} p\,(x,Q^2).
\end{equation}
From the theoretical point of view, there are two ways to avoid the 
problem of dealing with the unphysical region $x\rightarrow 0$. 
The first one is to work in $x$-space and obtain directly the 
evolution of parton distributions (not of their moments). 
The most known methods for solving the $Q^2$ evolution equations 
for parton distributions in $x$-space are brute-force \cite{J4}, 
Laguerre-polynomial \cite{J5} or Chebyshev-polynomial \cite{J6} 
approaches. In this way, the truncated moment can be simply found 
by integrating the $x$-space solutions $p(x,Q^2)$ over the cut range 
$x_0\leq x\leq 1$ (see Sect.~\ref{sec:2.1}).
An alternative way is use the evolution equations directly for 
truncated moments. The appropriate DGLAP evolution equations for 
the truncated moments have been derived in \cite{J2}. Authors have 
shown that these equations, though not diagonal, can be solved 
with a quite good precision for $n\geq 2$, even for a very small 
number of terms in the expansion series. In a case of the first 
moment, the accuracy is worse and more terms in the $G_n$ expansion 
must be taken into account. We briefly recall the idea of solving 
the evolution equations for truncated moments in Sect.~\ref{sec:2.2}.
Basing on this idea, we have derived the evolution equation for 
truncated first moment in a diagonal form. The appropriate 
integro-differential equation contains only one function - 
$q_1(x,Q^2)$, which denotes truncated at $x$ first moment, without 
coupling to the other, higher moments. Then, using the full Mellin 
moments approach, we have found the small-$x=x_0$ behaviour of the 
function $q_1(x,Q^2)$, which is simply the truncated at low-$x_0$ 
first moment. Detailed description is given in Sect.~\ref{sec:2.3}.

\subsection{LO DGLAP evolution equations with use of the Chebyshev-
polynomial expansion.}
\label{sec:2.1}
The Chebyshev polynomials technique \cite{J7} was successfully 
used by J.Kwieci\'nski in many QCD treatments e.g. \cite{J6}.
Using this method one obtains the system of linear differential 
equations instead of the original integro-differential ones. 
For example, in order to solve the LO DGLAP evolution equation 
for the nonsinglet parton distribution $p\equiv q^{NS}$:
\begin{equation}\label{eq.4}
\frac{\partial p\,(x,t)}{\partial t}=\frac{\alpha_s(t)}{2\pi}
\int\limits_x^1\frac{dz}{z} P_{qq}\left(\frac{x}{z}\right) p\,(z,t)
\end{equation}
one has to expand functions $p(x,t)$ into the series of the Chebyshev
polynomials. In this way, the integration over $z$ in the evolution 
equation (\ref{eq.4}) can be performed, what leads to the system of 
linear differential equations:
\begin{equation}\label{eq.5}
\frac{dp\,(x_i,t)}{dt} = \sum\limits_{j=0}^{N-1} H_{ij} p\,(x_j,t).
\end{equation}
This system can be solved by using the standard Runge-Kutta method with
initial conditions given by the input parametrisation $p(x_j,t_0)$.
Truncated at $x_0$ moments are simply computed numerically via the 
integrating (\ref{eq.3}).
The Chebyshev expansion provides a robust method of discretising 
a continuous problem. This allows computing the parton distributions 
for "not too singular" input parametrisation in the whole
$x\in (0;1)$ region. More detailed description of the Chebyshev 
polynomials method in the solving the QCD evolution equations is given 
e.g. in Appendix of \cite{J3}.

\subsection{Evolution equations for truncated moments.}
\label{sec:2.2}
LO DGLAP evolution equation for the truncated at $x_0$ Mellin moment of
the nonsinglet structure function found by Forte, Magnea, Piccione and
Ridolfi (FMPR) \cite{J2} has a form:
\begin{equation}\label{eq.6}
\frac{\partial {\it M}\big[p\big](x_0,n,t)}{\partial t}=
\frac{\alpha_s(t)}{2\pi}
\int\limits_{x_0}^1 dy\: y^{n-1} p\,(y,t) G_{n}\left(\frac{x_0}{y}\right),
\end{equation}
where
\begin{equation}\label{eq.7}
t\equiv \ln\frac{Q^2}{\Lambda_{QCD}^2}
\end{equation}
and 
\begin{equation}\label{eq.8}
G_n(z)\equiv\int\limits_z^1 dy\: y^{n-1}P_{qq}(y)
\end{equation}
is the truncated anomalous dimension.
Expanding $G_n(x_0/y)$ into Taylor series around $y=1$ gives
\begin{equation}\label{eq.9}
\frac{\partial {\it M}\big[p\big](x_0,n,t)}{\partial t}=
\frac{\alpha_s(t)}{2\pi}
\sum\limits_{j=0}^m C_{jn}^{(m)}(x_0) \:{\it M}\big[p\big](x_0,n+j,t).
\end{equation}
Eq.(\ref{eq.9}) is not diagonal but each $n$-th moment couples only 
with ($n+j$)-th ($j\geq 0$) moments. Because the series of couplings 
to higher moments is convergent and furthermore the value of 
($n+j$)-th moments decreases rapidly in comparison to the $n$-th 
moment, one can retain from (\ref{eq.9}) the closed system of $m+1$ 
equations:
\begin{displaymath}
\frac{\partial {\it M}\big[p\big](x_0,n,t)}{\partial t}
=\frac{\alpha_s(t)}{2\pi}
\end{displaymath}
\begin{equation}\label{eq.10}
\times\:\sum\limits_{j=n}^{N_0+m} D_{nj}^{(N_0+m-n)}(x_0) 
\:{\it M}\big[p\big](x_0,j,t).
\end{equation}
Here 
\begin{equation}\label{eq.11}
N_0\leq n\leq N_0+m,
\end{equation}
where
$N_0$ is the lowest considered moment and $D$ is a triangular matrix.
The solution of (\ref{eq.10}) has the form:
\begin{eqnarray}\label{eq.12}
&{\it M}\big[p\big](x_0,n,t) = 
\sum\limits_{k=n+1}^{N_0+m} A_{nk}(x_0){\it M}\big[p\big](x_0,k,t)&\nonumber\\
&+\big({\it M}\big[p\big](x_0,n,t_0) - 
\sum\limits_{k=n+1}^{N_0+m}
A_{nk}(x_0){\it M}\big[p\big](x_0,k,t_0)\big)&\nonumber\\
&\times\: \exp\left(c_f D_{nn}^{(m)}(x_0)\ln{\frac{t}{t_0}}\right).&
\end{eqnarray}
Matrix elements $D_{ij}^{(m)}(x_0)$ and $A_{ij}(x_0)$ are given in 
\cite{J2,J3}.
In \cite{J3} results (\ref{eq.12}) have been compared to those, 
obtained with use of the Chebyshev-polynomial technique. The agreement 
of both approaches is excellent for higher moments ($n\geq 2$) and not 
too large $x_0\leq 0.1$, even for a small number of terms ($m=4$) 
in the truncated series (\ref{eq.12}). It also does not depend 
strongly on the scale $Q^2$ or the input parametrisation. In a case 
of the truncated first moment, a similar accuracy requires more 
terms ($m\geq 30$) taken into account.\\
In the next section we estimate the small-$x_0$ behaviour of the 
truncated first moment ${\it M}\big[p\big](x_0,n=1,t)$.

\subsection{Small-$x_0$ behaviour of the truncated first moment.}
\label{sec:2.3}
We would like to present a possible way of an approximate determination 
the small-$x_0$ behaviour of the truncated first moment 
${\it M}\big[p\big](x_0,1,t)$. Our starting point is the evolution 
equation (\ref{eq.6}), which for the first moment has the form
\begin{equation}\label{eq.13}
\frac{\partial q_1(x_0,t)}{\partial t}=
\frac{\alpha_s(t)}{2\pi}
\int\limits_{x_0}^1 dy\: p\,(y,t) G_{1}\left(\frac{x_0}{y}\right).
\end{equation}
Here we denote in short the truncated first moment by $q_1(x_0,t)$:
\begin{equation}\label{eq.14}
q_j(x_0,t)\equiv {\it M}\big[p\big](x_0,j,Q^2).
\end{equation}
Inserting $G_1(z)$ in the LO approximation:
\begin{equation}\label{eq.15}
G_1(z) = \frac{8}{3}\ln (1-z) + \frac{4}{3}\left(z+\frac{z^2}{2}\right),
\end{equation}
we obtain
\begin{displaymath}
\frac{\partial q_1(x_0,t)}{\partial t}
=\frac{2\alpha_s(t)}{3\pi}[\: x_0\: q_0(x_0,t)
\end{displaymath}
\begin{equation}\label{eq.16}
+ \frac{x_0^2}{2}q_{-1}(x_0,t) 
- 2\sum\limits_{k=1}^{\infty}\frac{x_0^k}{k}q_{1-k}(x_0,t)\:].
\end{equation}
In the above formula we have used the expansion of\\ $\ln (1-z)$:
\begin{equation}\label{eq.17}
\ln (1-z) = - \sum\limits_{k=1}^{\infty}\frac{z^k}{k}.
\end{equation}
Taking into account that
\begin{equation}\label{eq.18}
q_j(x_0,t) = x_0^{j-1}q_1(x_0,t)
+ (j-1)\int\limits_{x_0}^{1} dy\: y^{j-2}q_1(y,t),
\end{equation}
we find
\begin{displaymath}
\frac{\partial q_1(x_0,t)}{\partial t} =
\frac{2\alpha_s(t)}{3\pi}\big[q_1(x_0,t)\big(\frac{3}{2}
-2\sum\limits_{k=1}^{\infty}\frac{1}{k}\big)
\end{displaymath}
\begin{equation}\label{eq.19}
+\int\limits_{x_0}^1 dy\: \big(2\sum\limits_{k=1}^{\infty}
\frac{x_0^k}{y^{k+1}} - \frac{x_0}{y^2} 
- \frac{x_0^2}{y^3}\big)q_1(y,t)\big].
\end{equation}
The above result is LO DGLAP evolution equation for the truncated 
first moment of the nonsinglet parton distribution. This formula 
contains only $q_1$ and there is no mixing between $q_1$ and higher 
or lower moments $q_j$. Therefore we can adopt to (\ref{eq.19}) well 
known approach and write the evolution equation of $q_1$ in the
moment space, what reads as follows:
\begin{equation}\label{eq.20}
\frac{\partial {\it M}\big[q_1\big](n,t)}{\partial t} = 
\frac{2\alpha_s(t)}{3\pi} H(n){\it M}\big[q_1\big](n,t).
\end{equation}
$H(n)$ is given by
\begin{equation}\label{eq.21}
H(n) = 2\sum\limits_{k=1}^{\infty}\left(\frac{1}{n+k} - \frac{1}{k}\right)
+ \frac{3}{2} - \frac{1}{n+1} - \frac{1}{n+2}.
\end{equation}
In this "moment of moment" approach (MM) we have obtained a simple equation 
for a some complicated structure ${\it M}\big[q_1\big](n,t)$, which is the 
(full) $n$-th moment of the truncated first moment of the parton density, 
namely
\begin{equation}\label{eq.22}
{\it M}\big[q_1\big](n,t) = \int\limits_{0}^{1} dx\:x^{n-1}
\int\limits_{x}^{1}dy\:p\,(y,t).
\end{equation}
The solution of (\ref{eq.20}) is
\begin{equation}\label{eq.23}
{\it M}\big[q_1\big](n,t) = {\it M}\big[q_1\big](n,t_0)\exp[a(t)H(n)],
\end{equation}
where ${\it M}\big[q_1\big](n,t_0)$ is a initial value of 
${\it M}\big[q_1\big]$ for the low scale $t_0$:
\begin{equation}\label{eq.24}
{\it M}\big[q_1\big](n,t_0) = \int\limits_{0}^{1} dx\:x^{n-1}
\int\limits_{x}^{1}dy\:p\,(y,t_0)
\end{equation}
and
\begin{equation}\label{eq.25}
a(t) = \frac{8}{33-2N_f}\ln\frac{t}{t_0}
\end{equation}
with $N_f$ - the number of the quark flavours.
Hence $q_1(x,t)$ is given by the inverse Mellin transform
\begin{equation}\label{eq.26}
q_1(x,t) = \frac{1}{2\pi i}\int\limits_{c-i\infty}^{c+i\infty}
dn\: x^{-n} {\it M}\big[q_1\big](n,t_0)\exp[a(t)H(n)].
\end{equation}
The behaviour of $q_1(x,t)$ for $x\rightarrow 0$ is governed by
the leading (i.e. rightmost) singularities of ${\it M}\big[q_1\big](n,t_0)$ 
as well as $H(n)$ in the $n$ complex plane \cite{J8}. If we parametrise 
the input parton distribution $p(x,t_0)$ in the general form
\begin{equation}\label{eq.27}
p\,(x,t_0) \sim x^{a_1}(1-x)^{a_2},
\end{equation}
we obtain
\begin{displaymath}
{\it M}\big[q_1\big](n,t_0) \sim \frac{1}{n}\:\beta (a_1+1,a_2+1)-
\sum\limits_{k=0}^{k_{max}}\frac{(-1)^k}{k!}
\end{displaymath}
\begin{equation}\label{eq.28}
\times\: \frac{\Gamma (a_2+1)}{\Gamma(a_2+1-k)(a_1+1+k)(n+a_1+1+k)}.
\end{equation}
$\Gamma(z)$, $\beta(z_1,z_2)$ in (\ref{eq.28}) are Euler functions 
and $k_{max}$ depends on $a_2$ in the following way:
\begin{equation}\label{eq.29}
k_{max} = \cases{\infty & $\textrm{for fractional}\; a_2 > 0$ \cr 
a_2 & $\textrm{for whole}\; a_2\geq 0$ \cr}
\end{equation}
One can see from (\ref{eq.21}) and (\ref{eq.28}) that $H(n)$ has the 
rightmost pole at $n=-1$, while ${\it M}\big[q_1\big](n,t_0)$ at $n=0$ 
and $n=-a_1-1$. In this way, for the nonsingular at small-$x$ shape of 
the starting distribution $p(x,t_0)$ ($a_1=0$), the simple pole at $n=0$ 
and the essential singularity at $n=-1$ are the leading ones. Then the 
small-$x_0$ behaviour of the truncated first moment can be determined 
by the method of steepest descent. We find
\begin{displaymath}
q_1(x,t) \approx \frac{1}{a_2+1} - \sqrt{\frac{e}{2\pi}}\:
\beta (a_2+1,z)\: x_0\: z^{1.5}[z+2a(t)]^{-0.5}
\end{displaymath}
\begin{equation}\label{eq.30}
\times\: exp\left(a(t)H(z-1)+0.5\sqrt{1-4a(t)\:ln(x_0)}\right),
\end{equation}
where
\begin{equation}\label{eq.31}
z\equiv -\frac{1+\sqrt{1-4a(t)\:ln(x_0)}}{2\: ln(x_0)}.
\end{equation}
If we consider more singular input parametrisation $p(x,t_0)$ ($a_1<0$), 
this singular small-$x$ behaviour remains stable against LO $Q^2$ QCD 
evolution. In this case the approximate behaviour of the truncated first
moment $q_1(x,t)$ is governed by the leading simple poles of 
${\it M}\big[q_1\big](n,t_0)$, situated at $n=0$ and $n=-a_1-1$:
\begin{displaymath}
q_1(x,t) \approx \beta(a_1+1,a_2+1) - 
\sum\limits_{k=0}^{k_{max}}\frac{(-1)^k}{k!}
\end{displaymath}
\begin{equation}\label{eq.32}
\times\: \frac{\Gamma (a_2+1)\:x^{a_1+1+k}}{\Gamma(a_2+1-k)\:(a_1+1+k)}
exp[a(t)H(-a_1-1-k)].
\end{equation}
In the next section we compare results (\ref{eq.30})-(\ref{eq.32}) 
with those, obtained within approaches, described in \ref{sec:2.1} and
\ref{sec:2.2}.

\section{Results for the truncated first moment within three different 
approaches.}
\label{sec:3}
In this section we present numerical results for the truncated first 
moment of the nonsinglet parton distribution. Predictions obtained 
with use of different methods are denoted via CHEB, FMPR or MM, according 
to the applied approach (see Sect.~\ref{sec:2.1}, \ref{sec:2.2}, 
\ref{sec:2.3} respectively).\\
Thus $q_1^{CHEB}(x_0,t)$ results from integrating
\begin{equation}\label{eq.33}
q_1^{CHEB}(x_0,t)=\int\limits_{x_0}^1 dx\: x^{n-1} p^{CHEB}(x,t),
\end{equation}
where $p^{CHEB}(x,t)$ is the solution of (\ref{eq.4})-(\ref{eq.5}), while
$q_1^{FMPR}(x_0,t)$ is implied by (\ref{eq.12}) and has the form
\begin{eqnarray}\label{eq.34}
&q_1^{FMPR}(x_0,t) = \sum\limits_{k=2}^{m+1} A_{1k}(x_0)q_k(x_0,t)&\nonumber\\
&+\left(q_1(x_0,t_0) - \sum\limits_{k=2}^{m+1} A_{1k}(x_0)q_k(x_0,t_0)\right)
&\nonumber\\
&\times\: exp\left(c_f D_{11}^{(m)}(x_0)\ln{\frac{t}{t_0}}\right),&
\end{eqnarray}
where $q_k(x_0,t)$ is defined by (\ref{eq.14}).\\
Analytical approximate solutions (\ref{eq.30})-(\ref{eq.32}), describing 
the low-$x_0$ behaviour of $q_1(x_0,t)$ within "moment of moment" 
($q_1^{MM}$) approach are compared with $q_1^{CHEB}$ and $q_1^{FMPR}$ 
results in Figs.1-6. We consider two values of the number of terms in 
the truncated series for the FMPR-m predictions ($m=4$, $m=30$). 
We show the results for the truncated at $x_0$ first moment of $g_1^{NS}$,
which is a contribution to the Bjorken Sum Rule \cite{J1a}. In presented 
figures the total Bjorken Sum Rule is normalized to 1. We use two different
inputs at $Q_0^2=1$ ${\rm GeV}^2$, namely:
\begin{eqnarray}\label{eq.35}
p\,(x,Q_0^2)\equiv g_1^{NS}(x,Q_0^2) &\sim&  (1-x)^3, \\ \label{eq.36}
p\,(x,Q_0^2)\equiv g_1^{NS}(x,Q_0^2) &\sim&  x^{-0.4}(1-x)^{2.5}.
\end{eqnarray}
More singular at small-$x$ parametrisation (\ref{eq.36}) incorporates the 
latest knowledge about the low-$x$ behaviour of the polarised structure 
functions \cite{J9}.
\begin{figure}
\begin{center}
\resizebox{0.7\textwidth}{!}{
\includegraphics{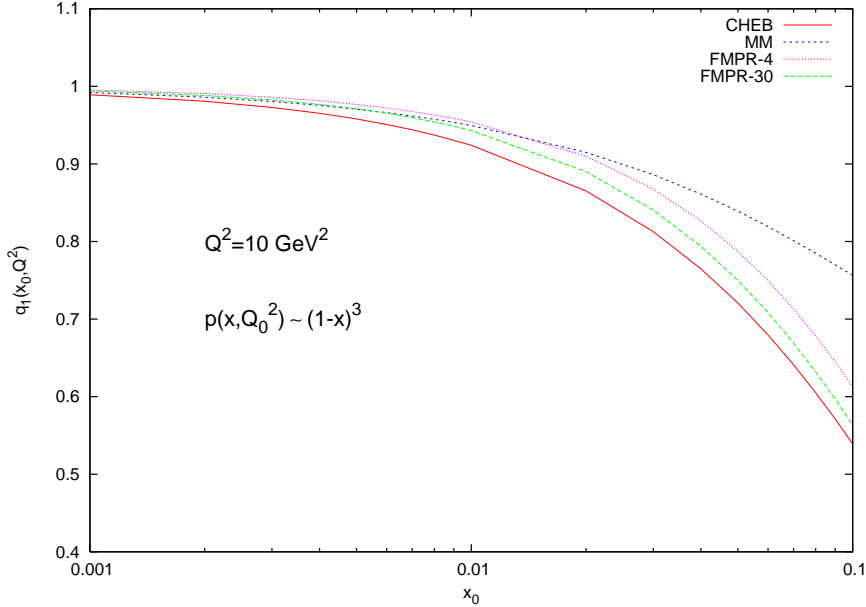}
}
\caption{Small $x_0$ behaviour of the truncated at $x_0$ first moment 
of the nonsinglet spin structure function $g_1^{NS}$ in a case of the 
flat input (\ref{eq.35}). A comparison of (\ref{eq.30}) (MM)
with the predictions based on the Chebyshev polynomials method (CHEB)
and FMPR approach (\ref{eq.34}) for two values of the number of terms 
in the truncated series: $m=4$, $m=30$ is shown. The Bjorken Sum Rule
is normalized to 1.}
\label{fig:1}
\end{center}
\end{figure}
\begin{figure}
\begin{center}
\resizebox{0.7\textwidth}{!}{
\includegraphics{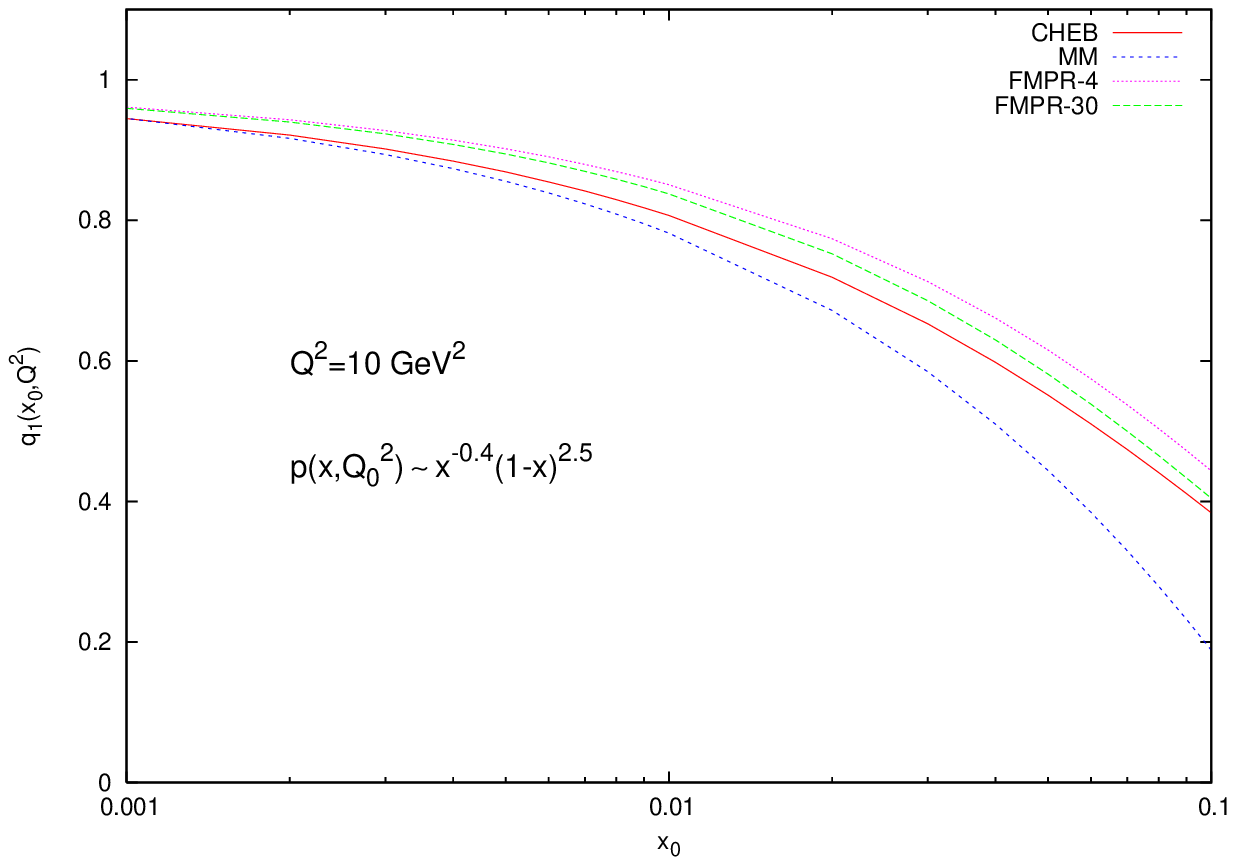}
}
\caption{Small $x_0$ behaviour of the truncated at $x_0$ first moment 
of the nonsinglet spin structure function $g_1^{NS}$ in a case of the 
steep input (\ref{eq.36}). A comparison of (\ref{eq.32}) (MM)
with the predictions based on the Chebyshev polynomials method (CHEB)
and FMPR approach (\ref{eq.34}) for two values of the number of terms 
in the truncated series: $m=4$, $m=30$ is shown. The Bjorken Sum Rule
is normalized to 1.}
\label{fig:2}
\end{center}
\end{figure}
\begin{figure}
\begin{center}
\resizebox{0.7\textwidth}{!}{
\includegraphics{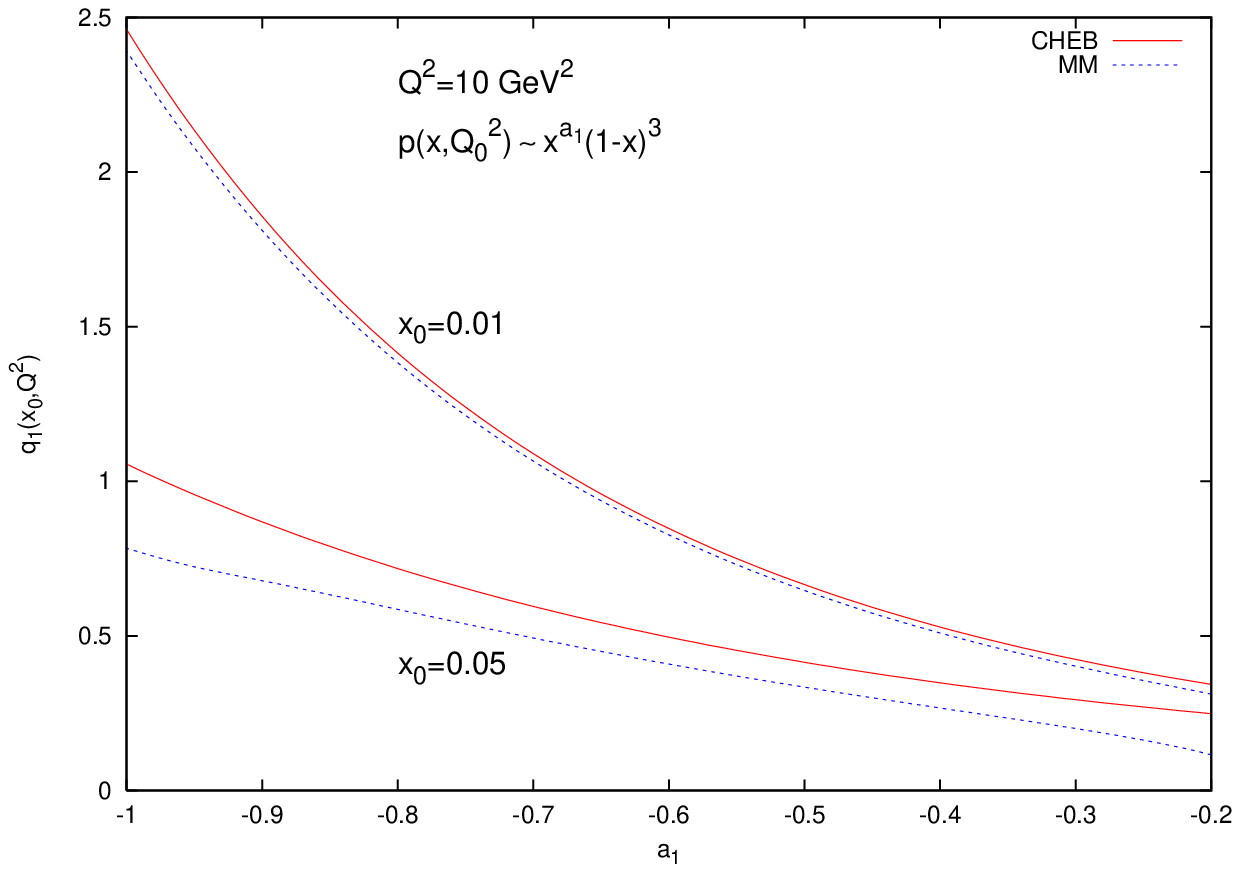}
}
\caption{Truncated first moment $q_1^{MM}$ (\ref{eq.32}) as a function 
of $a_1$ in the input parametrisation $p(x,Q_0^2)\sim x^{a_1}(1-x)^3$ for 
fixed $x_0=0.01$ and $x_0=0.05$. A comparison with the predictions based 
on the Chebyshev polynomials method is shown.}
\label{fig:3}
\end{center}
\end{figure}
\begin{figure}
\begin{center}
\resizebox{0.7\textwidth}{!}{
\includegraphics{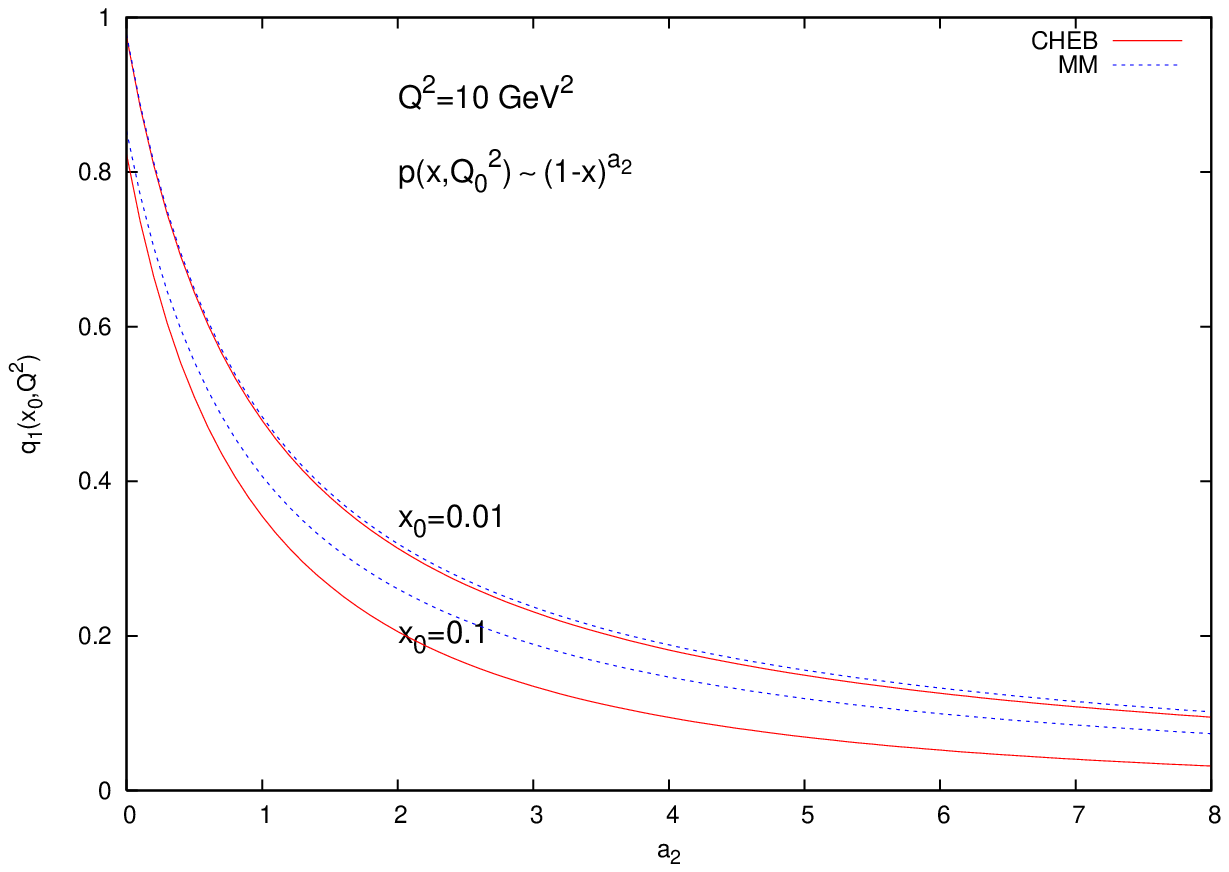}
}
\caption{Truncated first moment $q_1^{MM}$ (\ref{eq.30}) as a function 
of $a_2$ in the input parametrisation $p(x,Q_0^2)\sim (1-x)^{a_2}$ for 
fixed $x_0=0.01$ and $x_0=0.1$. A comparison with the predictions based 
on the Chebyshev polynomials method is shown.}
\label{fig:4}
\end{center}
\end{figure}
\begin{figure}
\begin{center}
\resizebox{0.7\textwidth}{!}{
\includegraphics{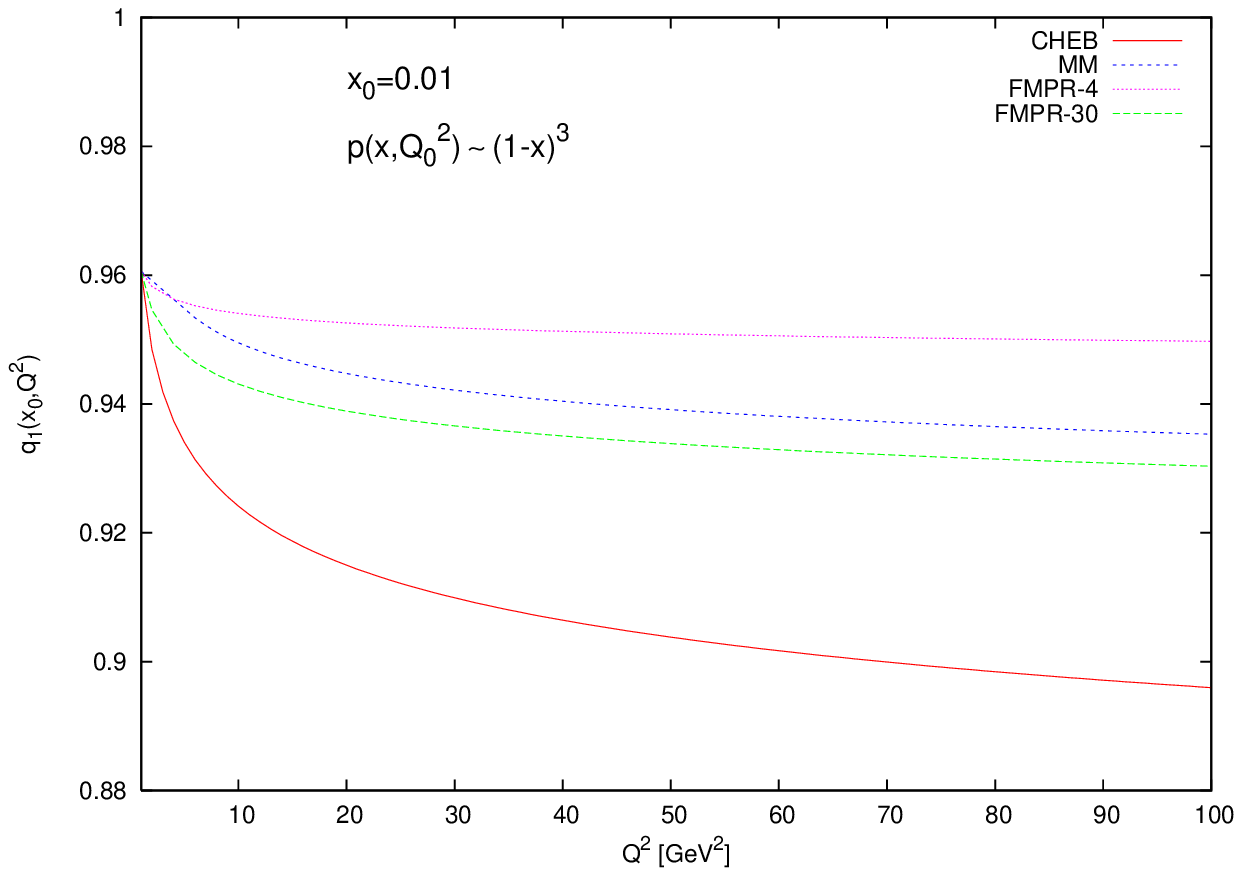}
}
\caption{$Q^2$ dependence of the truncated first moment $q_1^{MM}$ 
(\ref{eq.30}) at fixed $x_0=0.01$ for input parametrisation 
(\ref{eq.35}). A comparison to the predictions based on the Chebyshev 
polynomials method and FMPR approach (\ref{eq.34}) for two 
values of the number of terms in the truncated series: $m=4$, $m=30$ 
is shown.}
\label{fig:5}
\end{center}
\end{figure}
\begin{figure}
\begin{center}
\resizebox{0.7\textwidth}{!}{
\includegraphics{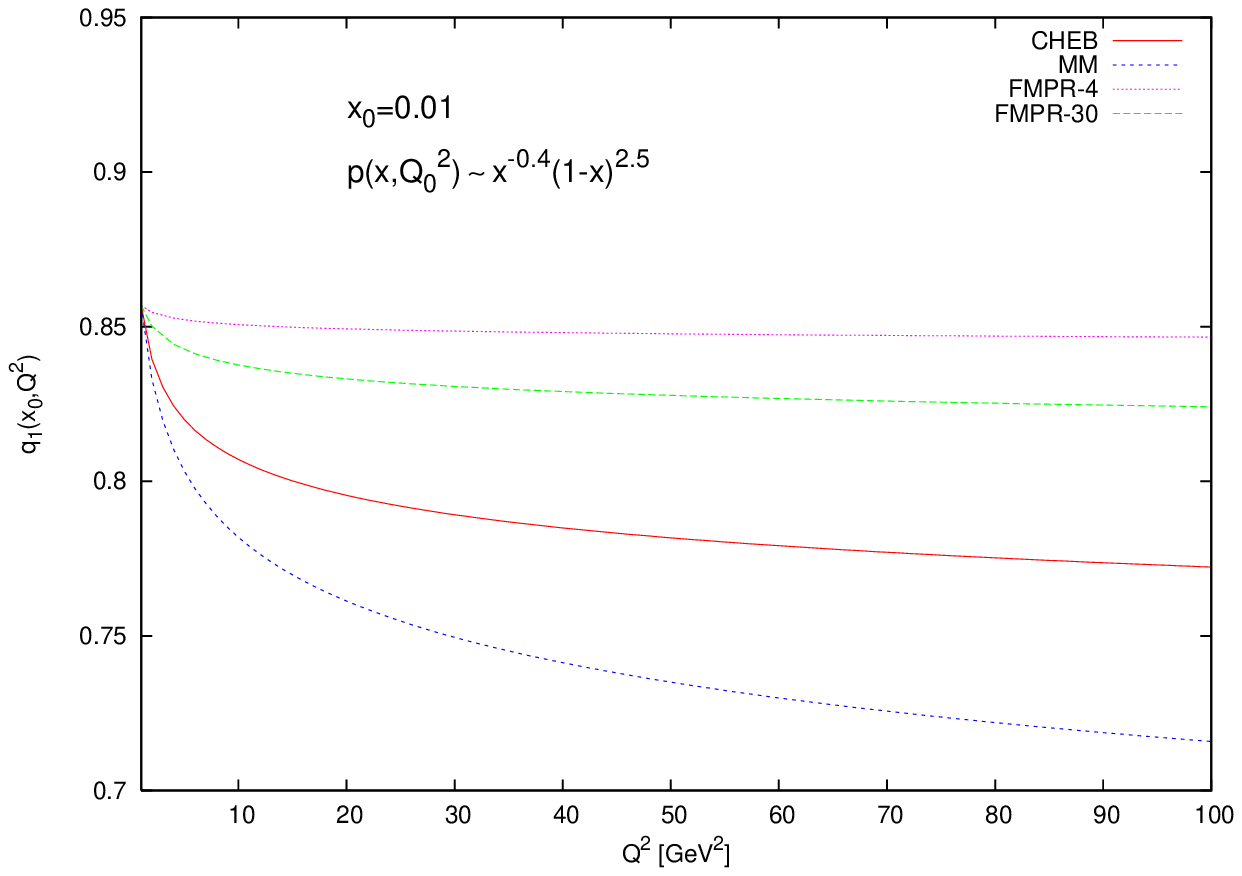}
}
\caption{$Q^2$ dependence of the truncated first moment $q_1^{MM}$ 
(\ref{eq.32}) at fixed $x_0=0.01$ for input parametrisation 
(\ref{eq.36}). A comparison to the predictions based on the Chebyshev 
polynomials method and FMPR approach (\ref{eq.34}) for two 
values of the number of terms in the truncated series: $m=4$, $m=30$ 
is shown.}
\label{fig:6}
\end{center}
\end{figure}
Truncated at $x_0=0.01$ integral $\int dx g_1^{NS}$ is reduced by about 
$8\%$ for the Regge input (\ref{eq.35}) and about $20\%$ for (\ref{eq.36})
in comparison to the total Bjorken Sum Rule.
Figures 1-6 show that for small-$x_0\leq 0.01$ there is a good agreement 
between the MM results (\ref{eq.30})-(\ref{eq.32}) and the predictions, 
obtained with use of the Chebyshev polynomial approach, which can be 
regarded as reliable. The percentage error 
\begin{equation}\label{eq.37}
E^{a}(x_0,Q^2) =
\frac{\mid q_1^{a}(x_0,Q^2)-q_1^{CHEB}(x_0,Q^2)\mid}
{q_1^{CHEB}(x_0,Q^2)}\cdot 100\%,
\end{equation}
where $a$ denotes MM, FMPR-4 or FMPR-30 results, is about $3\%$ in 
the case of MM solutions for $x_0=0.01$ and $Q^2=10$ ${\rm GeV}^2$, 
independently on the shape of the input parametrisation. Similar 
accuracy give taking into account more terms ($m=30$) in the truncated 
series within FMPR approach (FMPR-30), while for $m=4$ the error 
(\ref{eq.37}) is about $4\%$ in the case of the flat input and $6\%$ 
in the case of the more singular one. $E^{a}(x_0,Q^2)$ grows slowly 
with increasing $Q^2$ (see Figs. 5-6) and for $Q^2=100$ ${\rm GeV}^2$ 
we get: $E^{MM}(x_0=0.01,Q^2)\approx E^{FMPR-30}(x_0=0.01,Q^2)\approx 4\%$ 
for the flat parametrisation and $7\%$ respectively for the more singular 
input. Note, that for the truncation points $x_0\leq 0.01$ our 
approximate analytical solutions (\ref{eq.30})-(\ref{eq.32}) are as 
reliable as the $FMPR-30$ predictions and more exact than the $FMPR-4$ 
results. This does not depend either on the shape of the input 
parametrisation nor on the value of $Q^2$.\\
In the next section we determine the low-$x$ contribution to the Bjorken 
Sum Rule.

\section{Low-$x$ contribution to the Bjorken Sum Rule.}
\label{sec:4}
Among all moments of structure functions, the Bjorken Sum Rule (BSR) 
\cite{J1a} is one of the convenient tests of QCD. BSR is a fundamental
relation for polarised scattering, describing a relationship between spin
dependent DIS and the weak coupling constant defined in neutron $\beta$-decay.
In the limit of the infinite momentum transfer $Q^2$, the BSR reads:
\begin{equation}\label{eq.38}
I_{BSR} \equiv \Gamma_1^p - \Gamma_1^n =  
\int\limits_{0}^{1} dx\: (g_1^p(x)-g_1^n(x)) = \frac {1}{6}{\frac{g_A}{g_V}},
\end{equation}
where $g_V$ and $g_A$ are the vector and axial vector couplings. From recent
measurements \cite{JPDG} $g_A/g_V=1.2695\pm 0.0029$. BSR refers to the first
moment of the nonsinglet spin dependent structure function $g_1^{NS}$:
\begin{equation}\label{eq.39}
g_1^{NS}(x,Q^2) =  g_1^{p}(x,Q^2) - g_1^{n}(x,Q^2),
\end{equation}
where $g_1^p$ and $g_1^n$ are spin structure functions for proton and neutron.
The asymptotic relation (\ref{eq.38}) at finite $Q^2\gg\Lambda_{QCD}^2$ takes
a form with pQCD corrections:
\begin{displaymath}
\int\limits_{0}^{1} dx\: g_1^{NS}(x,Q^2) = \frac {1}{6}{\frac{g_A}{g_V}}
\end{displaymath}
\begin{equation}\label{eq.40}
\times\left[ 1 - \frac{\alpha_s}{\pi} - 3.583\left(\frac{\alpha_s}{\pi}
\right)^2 - 20.215\left(\frac{\alpha_s}{\pi}\right)^3\right].
\end{equation}
The validity of the sum rule is confirmed in polarised DIS at the level of
$10\%$ \cite{JSMC,JSLAC}. Ewaluation of the sum rules requires knowledge
of polarised structure functions over the entire region of $x$: 
$0\leq x\leq 1$. The experimentally accessible $x$ range for the spin dependent
DIS is however limited ($0.7>x>0.003$ for SMC data \cite{JSMC}, 
$0.6>x>0.023$ for HERMES data \cite{JHERMES}) and therefore
one should extrapolate results to $x=0$ and $x=1$. The extrapolation to
$x\rightarrow 0$, where structure functions grow strongly, is much more
important than that to $x\rightarrow 1$, where structure
functions vanish. The extrapolation towards $x=0$ suffers from large 
uncertainties, being essentially dependent on the used QCD fit. "Flexibility"
of the chosen parametrisation appears in the agreement with the experimental 
data, giving however enough freedom in the unmeasured regions 
\cite{JHERMES2005}. In a case of the BSR this allows for a significant 
reduction of the low-$x$ contribution.\\
In our approach we can test how the small-$x$ contribution to the BSR depends
on the different (less or more steep) input parametrisations at the initial 
scale $Q_0^2={\rm GeV}^2$:
\begin{equation}\label{eq.41}
g_1^{NS}(x,Q_0^2) = \eta\: x^{a_1}(1-x)^{a_2}.
\end{equation}
Here $\eta$ is a normalization factor:
\begin{equation}\label{eq.42}
\eta = \frac{I_{BSR}}{\beta (a_1+1,a_2+1)}
\end{equation}
The exponent $a_1$ controls the behaviour of the structure function 
$g_1^{NS}$ as $x\rightarrow 0$ and the factor $(1-x)^{a_2}$ ensures
the vanishing of $g_1^{NS}$ at $x\rightarrow 1$.
The percentage contribution to the BSR, coming from small-$x$ region 
$0\leq x\leq x_0$ is defined as:
\begin{equation}\label{eq.43}
r(x_0,Q^2) = \frac{\int\limits_{0}^{x_0} dx\: g_1^{NS}(x,Q^2)}
{\int\limits_{0}^{1} dx\: g_1^{NS}(x,Q^2)}\cdot 100\%
\end{equation}
The ratio $r$ for $x_0=0.01$ varies from a few to tens percents for different 
configurations of $-1\leq a_1\leq 0$ and $0\leq a_2\leq 8$.
The $r$-distribution at $x_0=0.01$ is shown in Fig.7.
One can see that the small-$x$ contribution to the BSR grows with increasing
$a_2$ and decreasing $a_1$. For the Regge flat parametrisation (\ref{eq.35})
$r$ is at the level of $5-10 \%$, what is significantly different from the 
result based on the input (\ref{eq.36}), where $r\sim 20 \%$. From theoretical 
analyses it is known that the small-$x$ behaviour of the nonsinglet polarised 
structure function $g_1^{NS}$ is governed by the double logarithmic terms i.e. 
$(\alpha_s ln^2x)^n$ \cite{JDL1,JDL2,J9}. This leads to the singular at low-$x$
form of $g_1^{NS}$:
\begin{equation}\label{eq.44}
g_1^{NS}(x,Q^2) \sim x^{-\lambda}
\end{equation}
with $\lambda\approx 0.4$. LO DGLAP approach with use of the singular input 
(\ref{eq.36}) pretends the double logarithmic $ln^2x$ resummation. 
Low-$x$ experimental data \cite{JSMC,JHERMES} clearly confirm the rise of
$g_1^{NS}$ in this region. However, the errors on the present data are too
large to reliably support or contradict this $x^{-0.4}$ behaviour. "Freedom" 
in the initial parametrisation to satisfy the small-$x$ experimental 
extrapolation of the BSR is seen in Fig.8. The experimental data can be
satisfactorily reproduced by e.g. nonsingular as $x\rightarrow 0$ input
$\sim (1-x)^6$ and by e.g. the singular one $\sim x^{-0.5}(1-x)^1$ as well.
SMC and SLAC measurements  imply that $10-20\%$ of the BSR comes from $x$
values less than $0.01$ \cite{JSMC,JSLAC,JBASS}. Also HERMES data
\cite{JHERMES} enable to determine the low-$x$ contribution to the BSR at
$0.023$ between $20-40\%$.
Wide range of these estimations could be restricted by new spin data concerning
this mystery and interesting small-$x$ region.
\begin{figure}
\begin{center}
\resizebox{0.7\textwidth}{!}{
\includegraphics{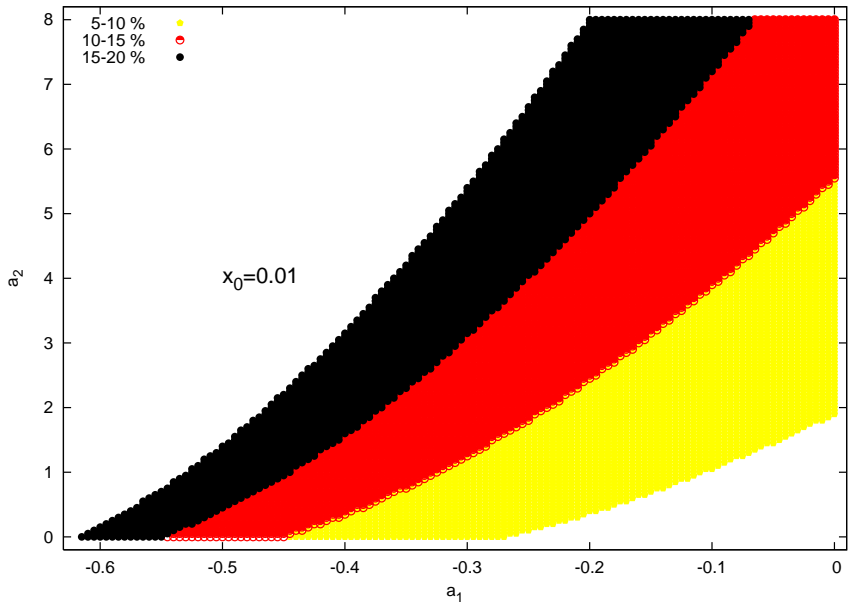}
}
\caption{Low-$x$ contribution to the Bjorken Sum Rule (\ref{eq.43})
for different $a_1$ and $a_2$ in the input parametrisation (\ref{eq.41}).
$x_0=0.01$ and $Q^2=5 {\rm GeV}^2$.}
\label{fig:7}
\end{center}
\end{figure}
\begin{figure}
\begin{center}
\resizebox{0.7\textwidth}{!}{
\includegraphics{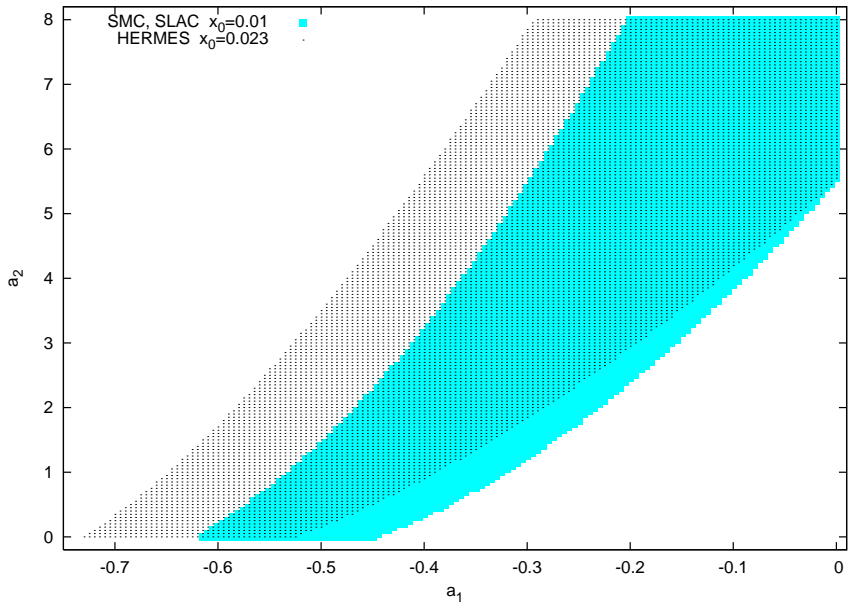}
}
\caption{Constraints on the parametrisation of $g_1^{NS}$, implied by the
experimental estimations of the low-$x$ contribution to the Bjorken Sum Rule.
SMC, SLAC: $10\%\leq r(x_0=0.01,5)\leq 20\%$ (full colour), HERMES:
$20\%\leq r(x_0=0.023,5)\leq 40\%$ (dotted).}
\label{fig:8}
\end{center}
\end{figure}

\section{Conclusions.}
\label{sec:5}
In this paper we have compared results for the truncated at $x_0$ first 
moment $q_1$ of the parton distribution obtained within different approaches. 
Thus we have solved numerically LO DGLAP evolution equation for the nonsinglet 
function in the $x$-space using Chebyshev polynomial expansion and then 
after integrating over $x$ we have got the prediction $q_1^{CHEB}$, 
which can be treated as an exact one. Next, using evolution equations 
written in the moment space, we have found the closed system of
$m+1$ solutions for truncated moments, where obtained $q_1^{FMPR-m}$ result
is expressed by values of the $m$ higher moments. Considered number of terms 
in the truncated series was $m=4$ and $m=30$. Working in the moment space
we have also found an alternative way to determine the small-$x_0$ behaviour
of the truncated first moment. Taking into account the relation between $n$-th
and $j$-th truncated moment, we were able to derive the evolution equation, 
which does not contain mixing between different moments. Then, adopting the
standard analytical method of full moments to the case of the first 
truncated moment, we have found approximate behaviour of $q_1$ as 
$x\rightarrow 0$. In this way the inverse Mellin transform performed with use 
of the method of steepest descent implies the result $q_1^{MM}$ within our 
modified "moment of moment" approach. 

We have shown that for small-$x_0\leq 0.01$ there is a good agreement 
between the MM results and the reliable predictions, obtained with use of 
the Chebyshev polynomial method. This agreement occurs independently 
either on the shape of the input parametrisation or on the value of $Q^2$.
It has been also found, that for small $x_0$ the accuracy of $q_1^{MM}$ 
and $q_1^{FMPR-30}$ results are similar, being clearly better than in
the case of $q_1^{FMPR-4}$ predictions.\\
We have presented results concerning the spin structure function $g_1^{NS}$
and the truncated at $x_0=0.01$ contribution to the Bjorken Sum Rule. It 
has been found, that the choice of the input parametrisation has a large
impact on the evaluation of the low-$x$ contribution to the BSR. This
contribution can vary from a few percents for the flat ($\sim const$) input to
tens percents for the steep ($\sim x^{-0.5}$) one. Recent experimental data
confirm the rise of the polarised structure functions at small-$x$. However, 
because of the large uncertainties, reliable support or contradiction of the
theoretical expectations is still out of reach.

\end{document}